\documentclass[lettersize,journal]{IEEEtran}
\usepackage{amsmath,amsfonts}
\usepackage{algorithmic}
\usepackage{algorithm}
\usepackage{array}
\usepackage[caption=false,font=normalsize,labelfont=sf,textfont=sf]{subfig}
\usepackage{textcomp}
\usepackage{stfloats}
\usepackage{url}
\usepackage{verbatim}
\usepackage{graphicx}
\usepackage{amssymb}
\usepackage{cite}

\usepackage{color}

\begin{document}

\title{Extended $k$-$u$ Fading Model in mmWave Communication: Statistical Properties and Performance Evaluations}

\author{Jiahuan Wu, Xiao-Ping Zhang,~\IEEEmembership{Fellow,~IEEE} , Xinchun Yu,~\IEEEmembership{Member,~IEEE}, Yuhan Dong,~\IEEEmembership{Senior Member,~IEEE}
\thanks{This work is supported by the National Natural Science Foundation of China under Grant 62388102, by the Shenzhen Ubiquitous Data Enabling Key Lab under Grant ZDSYS20220527171406015, and by the Tsinghua Shenzhen International Graduate School-Shenzhen Pengrui Endowed Professorship Scheme of Shenzhen Pengrui Foundation.\textit{ (Corresponding author: Xiao-Ping Zhang)}}
\thanks{Jiahuan Wu, Xiao-Ping Zhang, and Yuhan Dong are with the
Shenzhen Key Laboratory of Ubiquitous Data Enabling and  Shenzhen International Graduate School, Tsinghua University, Shenzhen 518055, China (e-mail: wujiahua24@mails.tsinghua.edu.cn; xpzhang@ieee.org; dongyuhan@sz.tsinghua.edu.cn)}
\thanks{Xinchun Yu is with the Zhejiang Key Laboratory of Big Data and Future ECommerce Technology, School of Computer Science and Technology, Zhejiang Gongshang University, Hangzhou 310018, China. (e-mail: yxc@mail.zjgsu.edu.cn)}
}



\maketitle

\begin{abstract}
In this paper, we present a novel small-scale fading model, named the extended $k$-$u$ model, which incorporates the imbalance of multipath clusters by adding a new parameter based on the original $k$-$u$ model. The extended $k$-$u$ model has more accurate modeling capability than the extended $\eta$-$u$ model in scenarios with line-of-sight (LoS) paths. Additionally, it is mathematically more tractable than the $a$-$k$-$\eta$-$u$ model. The extended $k$-$u$ model provides an effective channel modeling tool for millimeter (mmWave) LoS scenarios. Through theoretical derivations, we obtain closed-form expressions for the key statistical characteristics of this model, including the probability density function, the cumulative distribution function, moments of arbitrary order, and the moment generating function. Based on these statistics, this study further derives and analyzes the expressions for some performance metrics of the communication system, including the amount of fading, the probability of outage, the average bit error rate, and the effective rate. Using the measured fading data extracted from literature, which cover communication scenarios at 28 GHz, 65 GHz, and 92.5645 GHz with LoS paths, we apply the proposed model in mmWave scenarios and compare it with the $k$-$u$ model and the extended $\eta$-$u$ model. The results show that the extended $k$-$u$ model has better capability in characterizing such fading than the other two models, verifying that this extension enhances its ability to model LoS mmWave scenarios.
\end{abstract}


\section{Introduction}

\IEEEPARstart{W}{ireless} communication relies on the transmission of electromagnetic signals between transceivers, where the quality of signal propagation directly dictates the reliability and efficiency of the entire communication system. When signals propagate through the environment, factors such as diffraction, scattering, and reflection trigger the superposition of multipath components at the receiver. Such a superposition not only gives rise to amplitude fluctuations, but also undermines the stability and reliability of reception performance. To better characterize these fading effects, researchers have developed a range of statistical models rooted in the stochastic modeling of communication environments. Among classical small-scale fading models, the Rayleigh, Rician, and Nakagami models are the most widely adopted~\cite{b1,b2,b3}.

With the continuous advancement of communication technologies, application scenarios have grown increasingly diverse. Communication environments have expanded beyond traditional outdoor channels to encompass emerging scenarios such as indoor spaces, underwater communication systems, body area networks (BANs), vehicular networks, and the Internet of Things (IoT). Concurrently, signal frequencies are shifting toward higher bands, including millimeter-wave (mmWave) and terahertz (THz) bands. In such complex and emerging scenarios, conventional fading models often struggle to accurately characterize channel characteristics, leading to potential deviations in performance evaluation. For example, studies~\cite{b4} have reported that in various mmWave environments, classical fading models cannot adequately capture the actual fading behavior of wireless channels. As a foundational component of channel modeling, fading models are critical to performance analysis for fifth-generation (5G) and next-generation wireless communication systems \cite{b5}. Inaccurate channel modeling not only undermines the reliability of system evaluation but also impedes the rational design of communication systems—including key modules such as signal detection, coding, and resource allocation. Thus, developing more accurate and generalized fading models capable of adapting to diverse communication scenarios remains a prominent and pressing research challenge in wireless communications.

\subsection{Prior Works}
Reference~\cite{b6} introduced the concept of multipath clusters in Rayleigh fading and proposed a more general $\eta$-$u$ model. 
In~\cite{b7}, Rician fading was extended to clustered multipath environments, leading to the $k$-$u$ model. The $k$-$u$ model is more suitable for scenarios with a line-of-sight (LoS) component, while the $\eta$-$u$ model is more appropriate for non-line-of-sight (NLoS) cases. Experimental validation in outdoor 500~MHz, indoor 1.8~GHz, and indoor 10~GHz scenarios~\cite{b8} demonstrated that the $k$-$u$ and $\eta$-$u$ models outperform classical models in characterizing small-scale fading. These models have been widely applied in the design and performance analysis of communication systems~\cite{b9,b10,b11,b12,b13}. Reference~\cite{b14} generalized the uniform scattering assumption of Rayleigh fading to non-uniform environments and proposed the $\alpha$-$u$ model. In~\cite{b15}, it was further shown that the $\alpha$-$u$ model provides more accurate channel modeling for vehicular communications. 
Building upon this, environmental non-uniformity is incorporated into the $k$-$u$ and $\eta$-$u$ models in~\cite{b16}, yielding the $\alpha$-$k$-$u$ and $\alpha$-$\eta$-$u$ models. Due to their enhanced generality, these models have attracted significant attention~\cite{b17,b18,b19}.

By introducing concepts such as multipath clusters and environmental inhomogeneity, the aforementioned models can address the issue that traditional models fail to capture the real characteristics of channels in certain scenarios. In mmWave communication, multipath is relatively sparse~\cite{b20}. In scenarios with abundant multipaths, the imbalance in the number of multipaths between the in-phase and quadrature components of the received signal is negligible. But this imbalance is increasingly amplified and becomes non-negligible in mmWave scenarios. The aforementioned models do not account for this problem, resulting in their poor performance on mmWave channels~\cite{b21}. The models that take into account the imbalance of multipath clusters are ~\cite{b22,b23,b24}. Reference~\cite{b22} integrates the $a$-$k$-$u$ and $a$-$n$-$u$ models, incorporates the multipath imbalance factor, and proposes the most general $a$-$k$-$n$-$u$ model. Reference~\cite{b25} has proved through experiments in the 25 -- 40 GHz mmWave bands that the $a$-$k$-$n$-$u$ model performs better than other models in the mmWave band. The disadvantage of the $a$-$k$-$n$-$u$ model is that it is mathematically intractable, as its probability density function (PDF) of the signal envelope and cumulative distribution function (CDF) of the signal envelope remain in integral form. This limitation hinders its direct use in communication system analysis and design. In~\cite{b23}, cluster imbalance was introduced into the $\eta$-$u$ model, leading to an extended $\eta$-$u$ model. The extended $\eta$-$u$ model is more mathematically tractable than the $a$-$k$-$n$-$u$ model, since its key statistical quantities can be expressed in closed form. Reference ~\cite{b24} further incorporated environmental non-uniformity into this framework, resulting in the extended $a$-$n$-$u$ model. ~\cite{b23} and ~\cite{b24} have already been applied in the design and analysis of certain communication systems~\cite{b26,b27,b28}. 

\subsection{Motivations \& Contributions}
However, since the $\eta$-$u$ model tends to exhibit favorable applicability to scenarios without LoS path, for cases where the LoS component exists, the extended $\eta$-$u$ model may not be the optimal choice, and there is room for improvement in its performance. In 5G and future communication technologies, as the communication frequency bands move toward mmWave and higher frequencies, wireless channels become increasingly sparse and the LoS path grows significantly in importance. Consequently, communication systems may become more dependent on the presence of the LoS path and the extended $\eta$-$u$ model is not suitable for characterizing such scenarios. 

To provide a better channel modeling tool for 5G and future communication scenarios with LoS paths, this study leverages the suitability of the $k$-$u$ model for LoS scenarios, introduces the imbalance of multipath clusters into this model, and proposes an extended $k$-$u$ model. The extended $k$-$u$ model is more capable of characterizing fading in LoS scenarios than the extended $\eta$-$u$ model. Compared to the original $k$-$u$ model, the extended $k$-$u$ model incorporates an additional parameter from a mathematical perspective, which enhances its flexibility and generality to cover more fading scenarios. From a physical perspective, the multipath cluster imbalance introduced in the extended $k$-$u$ model enables it to outperform the original $k$-$u$ model in modeling mmWave communication scenarios. In addition, the proposed model exhibits better mathematical tractability than the $a$-$k$-$n$-$u$ model, and its basic statistical quantities can be expressed in closed-form. The contributions of this work are summarized as follows:
\begin{itemize}
\item We present an extended $k$-$u$ small-scale fading model, an effective channel modeling tool for mmWave LoS scenarios. This model addresses the lack of applicable tools in such scenarios, where the extended $\eta$-$u$ model is less applicable while the $a$-$k$-$\eta$-$u$ model faces the challenge of complex mathematical tractability.
\item Closed-form expressions for key statistics are derived, including the probability density function, the cumulative distribution function, moments of arbitrary order and the moment generating function.
\item Based on the extended $k$-$u$ model, the performance metrics of communication systems are analyzed, including the amount of fading, the outage probability, the average bit error rate (ABER) and the effective rate.
\item We apply the proposed model in mmWave scenarios, including 28 GHz, 65 GHz, and 92.5645 GHz bands and show that its capability to characterize mmWave LoS scenarios is superior to the $k$-$u$ model and the extended $\eta$-$u$ model.
\end{itemize}

\subsection{Organization}
The organization of the remaining parts of this paper is as follows. In Section \uppercase\expandafter{\romannumeral 2}, we present the mathematical formulation, physical significance, and advantages of the extended $k$-$u$ model. In Section \uppercase\expandafter{\romannumeral 3}, we present the detailed derivation process for the key statistical properties of the extended $k$-$u$ model.
In Section \uppercase\expandafter{\romannumeral 4}, we apply the extended $k$-$u$ model to derive and analyze the performance indicators of communication systems. In Section \uppercase\expandafter{\romannumeral 5}, we investigate the application of the extended $k$-$u$ model in the mmWave Los scenarios, and compares the performance of the extended $k$-$u$ model, the $k$-$u$ model and the extended $\eta$-$u$ model. Finally, In Section \uppercase\expandafter{\romannumeral 6}, we summarize the entire paper.

\section{Extended $k$-$u$ Model}
The mathematical model of the signal envelope for the extended $k$-$u$ model is given as
\begin{equation}
\label{2}
R^2 = \sum_{i=1}^{u}(X_i + w_i)^2 + \sum_{i=1}^{u \times p}(Y_i + q_i)^2.
\end{equation}
where $R$ denotes the envelope of the received signal, $u$ denotes the number of multipath clusters, and $p$ describes the imbalance of these multipath clusters. Each multipath cluster consists of an in-phase component and a quadrature component. Specifically, the in-phase component comprises $X_i$ and  $w_i$, while the quadrature component comprises $Y_i$ and $q_i$.  $X_i$ and $Y_i$ are independent Gaussian random variables, which represent the scattered wave components of the in-phase and quadrature components, respectively. They satisfy $\mathbb{E}[X_i] = \mathbb{E}[Y_i] = 0$ and $\mathbb{D}[X_i] = \mathbb{D}[Y_i] = \sigma^2$. In contrast, $w_i$ and $q_i$ are unknown constants, corresponding to the dominant components of the in-phase and quadrature components. Additionally, different multipath clusters are mutually independent.

The key difference between the extended $k$-$u$ model and the original $k$-$u$ model is that the former incorporates a parameter $p$ to characterize the imbalance of multipath clusters, and this introduction is necessary, as elaborated below. In the original $k$-$u$ model \cite{b8}, the number of multipath clusters in the in-phase and quadrature components is assumed identical, with both described by the same parameter $u$ . However, in practical scenarios, an imbalance may exist between their counts. When multipaths are abundant, this imbalance can be neglected. Under such conditions, the original $k$-$u$ model remains suitable for fading modeling. As signal frequencies rise to mmWave and higher bands, multipaths become relatively sparse ~\cite{b20}, and the influence of this imbalance is amplified to a non-negligible extent. Since the original $k$-$u$ model does not account for this imbalance, it will introduce significant errors into channel modeling for mmWave scenarios. Therefore, to improve the characterization capacity of the original $k$-$u$ model for mmWave scenarios, it is necessary to incorporate the imbalance of multipath clusters into the model framework. 

To minimize the parameter range and thereby simplify mathematical handling, we characterize the imbalance of multipath clusters by introducing the parameter $p$. The essence of multipath cluster imbalance lies in the inequality between the number of multipath clusters in the in-phase component $u$ and that in the quadrature component $u'$. This would result in two variables, both ranging from 0 to infinity. To address this, we retain the parameter $u$ for the component with more multipath clusters, while introducing $p$ to describe the component with fewer clusters. The range of $p$ is defined as  $0 \leq p \leq 1$ , ensuring the model can cover all possible imbalance scenarios. When $p=1$, the extended $k$-$u$ model degenerates into $k$-$u$ model. 

The extended $k$-$u$ offers the following advantages. Leveraging the inherent advantages of the original $k$-$u$ model, the extended $k$-$u$ model retains parameters $w_i$ and $q_i$ for characterizing the LoS component. This feature is absent from the extended $\eta$-$u$ model. Thus, the extended $k$-$u$ model tends to be more suitable for LoS scenarios than the extended $\eta$-$u$ model. Additionally, despite the introduction of an extra parameter, the extended $k$-$u$ model remains mathematically simpler than the $a$-$k$-$\eta$-$u$ model. Importantly, the key statistical properties of the extended $k$-$u$ model can still be expressed in closed-form, as will be shown in Section \uppercase\expandafter{\romannumeral 3}. The extended $k$-$u$ model provide an effective channel modeling tool for mmWave LoS scenarios. 

\section{Derivation of the Key Statistical Characteristics}
This section consists of a detailed derivation of several key statistical properties of the extended $k$-$u$ model, including the probability density function, the cumulative distribution function, and the moments of arbitrary order.
\subsection{Probability Density Function}
To obtain the PDF of the signal enevelope, let $A_i=(X_i + w_i)^2$ and $B_i=(Y_i + q_i)^2$, (\ref{2}) becomes
\begin{equation}
\label{3}
R^2 = \sum_{i=1}^{u}A_i + \sum_{i=1}^{u \times p}B_i.
\end{equation}

Through standard procedures, the PDF of the variable $A_i$ denoted by $f_{A_i}(a_i)$ can be derived as
\begin{equation}
\label{4}
f_{A_i}(x)=\frac{1}{\sqrt{2\pi\sigma^2x}}e^{-\frac{x+w_i^2}{2\sigma^2}}\cosh\left(\frac{w_i\sqrt{x}}{\sigma^2}\right),
\end{equation}
where $\cosh(x)$ denotes the hyperbolic cosine function ~\cite[eq. (1.311.3)]{b29}.

Applying the Laplace transform to $f_{A_i} (a_i)$ and using the Laplace transform pair \cite[eq. (3.546.2)]{b30}, we obtain
\begin{equation}
\mathcal{L}[f_{A_i}(x)] = \frac{1}{\sqrt{1 + 2s\sigma^2}} e^{-\frac{s w_i^2}{1 + 2s\sigma^2} },
\end{equation}
where, $s$ is denotes the Laplace variable.

Since $A_i$ and $B_i$ are independent random variables, the Laplace transform of the PDF of $R^2$ denoted by $f_{R^2}(r)$ is the product of the respective Laplace transforms of $A_i$ and $B_i$. The expression of the Laplace transform of $f_{R^2}(r)$ can be obtained as
\begin{align}
\mathcal{L}[f_{R^2}(x)] &= \left( \prod_{i=1}^{u} \mathcal{L}[f_{A_i}(x)] \right) \left( \prod_{i=1}^{u \times p} \mathcal{L}[f_{B_i}(x)] \right) \notag \\ 
&= \frac{1}{(1 + 2s\sigma^2)^{\frac{u + p}{2}}} e^{-\frac{s d^2}{1 + 2s\sigma^2} }, 
\end{align}
where $d^2 = \sum_{i=1}^{u} w_i^2 + \sum_{i=1}^{u} q_i^2$.

By applying the inverse Laplace transform to $\mathcal{L}[f_{R^{2}}(x)]$ and using ~\cite[eq. (2.2.2.1)]{b31}, $f_{R^2}(r)$ can be obtained as
\begin{equation}
\label{7}
f_{R^2}(r)=\frac{e^{-\frac{x+d^2}{2\sigma^2}}}{2\sigma^2}\left(\frac{\sqrt{x}}{d}\right)^{\frac{u(1+p)-2}{2}}I_{\frac{u(1+p)-2}{2}}\left(\frac{d}{\sigma^2}\sqrt{x}\right),
\end{equation}
where $I_v(x)$ denotes the modified Bessel function of the first kind of order $v$ ~\cite[eq. (9.6.3)]{b29}.

Through standard procedures, the PDF of $R$ denoted by  $f_R(r)$ can be obtained as
\begin{equation}
\label{8}
f_R(r)=\frac{r}{\sigma^2}e^{-\frac{r^2+d^2}{2\sigma^2}}\left(\frac{r}{d}\right)^{\frac{u(1+p)-2}{2}}I_{\frac{u(1+p)-2}{2}}\left(\frac{rd}{\sigma^2}\right).
\end{equation}

Define $k=\frac{d^2}{u(1+p)\sigma^2}$ as the ratio of the dominant component power to the scattered wave power and the root mean square value of the envelope  $R$ as $\hat{r}=\sqrt{\mathbb{E}[R^{2}]}=\sqrt{u(1+p)\sigma^{2}+d^{2}}$. By substituting  $k$ and $R$  into (\ref{8}) and rearranging, the final expression of $f_R(r)$ can be obtained as
\begin{align}
\label{rpdf}
f_R(r) = &\frac{(1+k)^{\frac{u(1+p)+2}{4}} (1+p)u}
               {k^{\frac{u(1+p)-2}{4}} e^{\frac{(1+p)uk}{2}} \hat{r}} 
           e^{-\frac{(1+k)(1+p)u r^2}{2\hat{r}^2}} \notag \\
       &\times \left( \frac{r}{\hat{r}} \right)^{\frac{u(1+p)}{2}} 
           I_{\frac{u(1+p)-2}{2}} \left( 
           \frac{\sqrt{k(1+k)} (1+p)u r}{\hat{r}} \right).
\end{align}

Through standard procedures, the PDF of the normalized envelope $P=R/\hat{r}$ denoted by $f_P(\rho)$ can be obtained as
\begin{align}
\label{10}
f_P(\rho) = &\frac{(1+k)^{\frac{u(1+p)+2}{4}} (1+p)u}
               {k^{\frac{u(1+p)-2}{4}} e^{\frac{(1+p)uk}{2}}} 
           e^{-\frac{(1+k)(1+p)u\rho^2}{2}} \notag \\
          &\times \rho^{\frac{u(1+p)}{2}} 
           I_{\frac{u(1+p)-2}{2}} \left( \sqrt{k(1+k)} (1+p)u\rho \right).
\end{align}

It can be observed that the PDF of the normalized envelope for the extended $k$-$u$ model is expressible in closed form. When setting $p=1$, we can derive the expression of the original $k$-$u$ model \cite{b8}. Despite the introduction of an extra parameter, the PDF of the extended $k$-$u$ model is not significantly more complex than that of the original $k$-$u$ model. Consistent with the original model, the PDF of the extended $k$-$u$ model consists of exponential functions, power functions, and modified Bessel functions of the first kind, with differences only present in certain parameters of these functions.

\begin{figure}[htbp]
    \centering
    
    \subfloat[$k=0.1,u=2.$]{
        \includegraphics[width=0.45\textwidth]{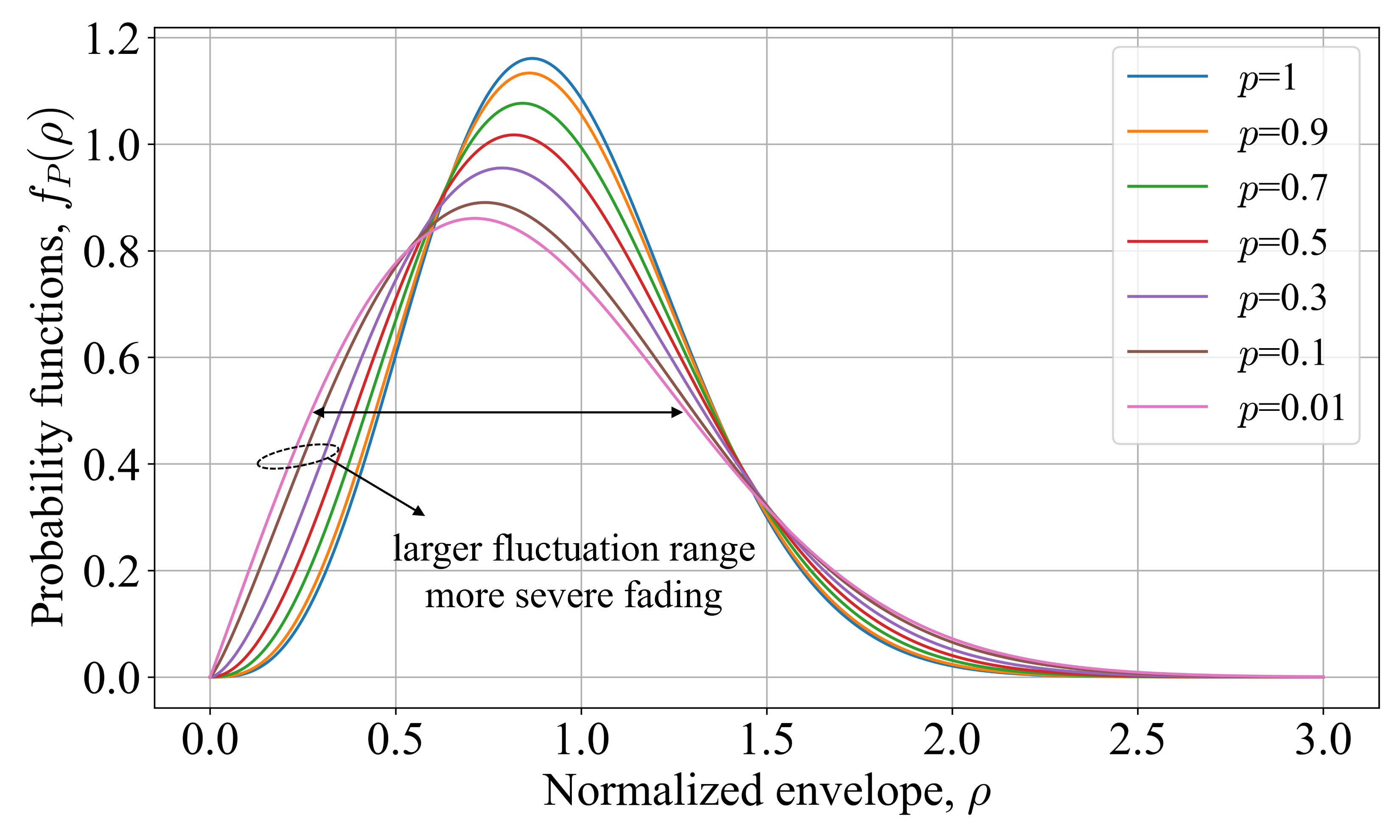} 
        \label{pdf1}
    }
    \hfill 
    
    \subfloat[$k=1,u=0.75.$]{
        \includegraphics[width=0.45\textwidth]{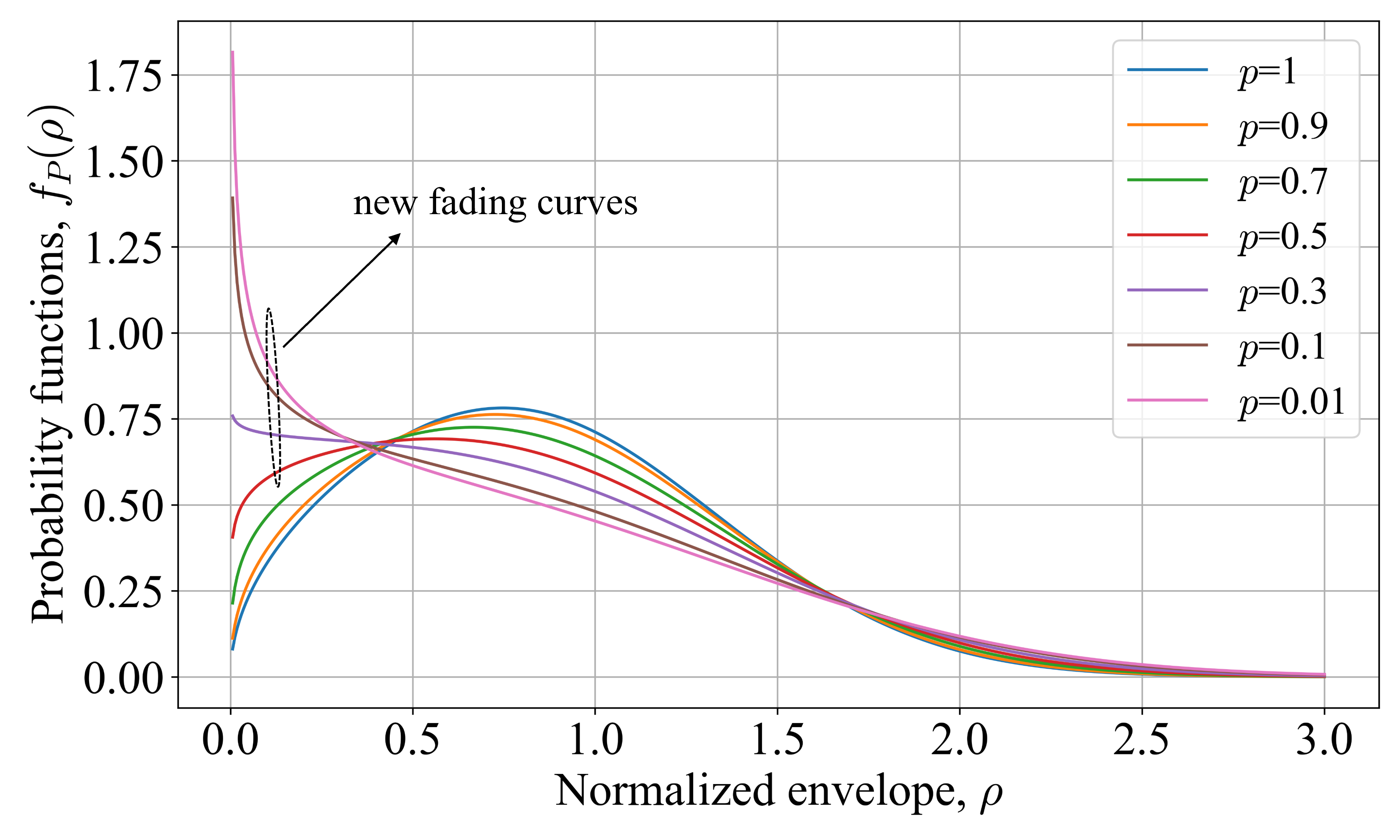} 
        \label{pdf2}
    }
    \hfill
    
    \subfloat[$k=5,u=0.65.$]{
        \includegraphics[width=0.45\textwidth]{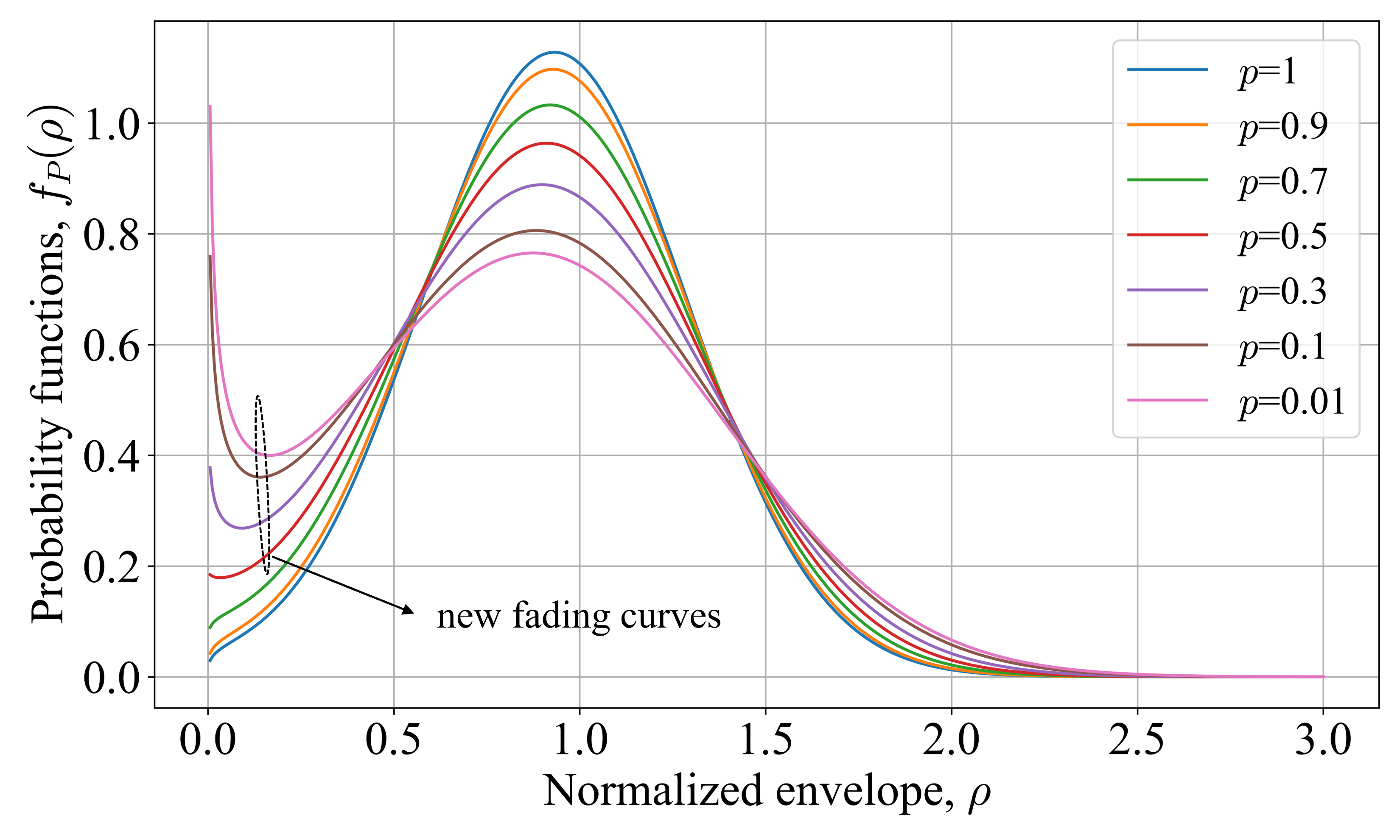} 
        \label{pdf3}
    }
    
    \caption{The probability function $f_P(\rho)$ of the extended $k$-$u$ model with different $p$. (Compared with the original $k$-$u$ model ($p=1$), the extended $k$-$u$ model offers enhanced coverage of fading scenarios, encompassing both novel and more severe fading conditions.)}
    \label{pdf}
\end{figure}

Fig. \ref{pdf} illustrates the PDFs of the normalized envelope for the extended $k$–$u$ model under different values of $p$. The curve corresponding to $p = 1$ also represents the original $k$–$u$ model. In Fig. \ref{pdf}(a), where the parameters are set to $k = 0.1$ and $u = 2$, it can be observed that as $p$ decreases, the peak of the PDF becomes broader, indicating that the signal fluctuation increases. Meanwhile, the peak gradually shifts away from $\rho = 1$, suggesting that the signal distribution deviates further from the original concentration. Both trends imply that the fading becomes more severe as $p$ decreases. Fig. \ref{pdf}(b) shows the PDFs for $k = 1$ and $u = 0.75$ under different values of $p$. As shown in the figure, with decreasing $p$. The shape of the PDF changes significantly. The peak gradually flattens and eventually disappears, resulting in a new curve profile. This behavior corresponds to a new fading scenario in which the signal envelope is no longer concentrated around $\rho = 1$, but rather spreads uniformly across different envelope amplitudes or even concentrates near $\rho = 0$, indicating a more severe fading condition. Fig. \ref{pdf}(c) presents the PDFs for $k = 5$ and $u = 0.65$ under different values of $p$. It can be seen that as $p$ decreases, the peak position remains almost unchanged, but its height gradually reduces. More importantly, a valley begins to appear in the curve, representing another distinct fading scenario. In this case, the majority of the signal remains concentrated near the original envelope level, while a portion of the signal power is distributed in deep fading regions.

\subsection{Cumulative Distribution Function}
The CDF of the normalized envelope $P$ denoted by $F_P(\rho)$ can be obtained through the following integral
\begin{align}
\label{31}
F_P(\rho) &= \int_0^\rho f_P(x)\mathrm{d}x = \frac{(1+k)^{\frac{u(1+p)+2}{4}} u(1+p)}{k^{\frac{u(1+p)-2}{4}} e^{\frac{(1+p)uk}{2}}} \notag \\
          &\quad \times \int_0^\rho e^{-\frac{(1+k)(1+p)u x^2}{2}} x^{\frac{u(1+p)}{2}} \notag \\
          &\quad \times I_{\frac{u(1+p)-2}{2}}\!\bigl(\sqrt{k(1+k)} u(1+p) x\bigr) \mathrm{d}x.
\end{align}

A variable substitution $x=\frac{t}{\sqrt{u(1+p)(1+k)}}$ is performed. After rearrangement, (\ref{31}) transforms into
\begin{align}
\label{32}
F_P(\rho) &= \frac{\bigl[(1+p)uk\bigr]^{-\frac{u(1+p)-2}{4}}}{e^{\frac{(1+p)uk}{2}}} \notag \\
          &\quad \times \int_0^{\sqrt{u(1+p)(1+k)}\rho} e^{-t^2/2} t^{\frac{u(1+p)}{2}} \notag \\
          &\quad \times I_{\frac{u(1+p)-2}{2}}\!\bigl(\sqrt{(1+p)uk}\,t\bigr) \mathrm{d}t.
\end{align}

To align with the form of the Marcum $Q$ function, (\ref{32}) is rearranged into the following expression
\begin{align}
\label{33}
F_P(\rho) &= 1 - \int_{\sqrt{u(1+p)(1+k)} \rho}^{\infty}
    \frac{1}{\left( \sqrt{(1+p)uk} \right)^{\frac{u(1+p)}{2} - 1}} \notag \\
    &\quad \times e^{-\frac{t^2 + \left( \sqrt{(1+p)uk} \right)^2}{2} }
    t^{\frac{u(1+p)}{2}} \notag \\
    &\quad \times I_{\frac{u(1+p)}{2} - 1} \left( \sqrt{(1+p)uk} \, t \right) \mathrm{d}t.
\end{align}

 The integral form of the Marcum $Q$ function \cite[eq. (4.41)]{b32} is given by
\begin{equation}
\label{34}
Q_M(a,b)=\int_b^\infty\frac{1}{a^{M-1}}e^{-\frac{t^2+a^2}{2}}t^nI_{M-1}(at) \mathrm{d}t.
\end{equation}

Utilizing the above integral form, the final expression of $F_P(\rho)$ can be obtained as
\begin{multline}
\label{11}
F_P(\rho) = 1 - Q_{\frac{u(1 + p)}{2}} \left( \sqrt{(1+p)uk}, \right. \\
\left. \sqrt{u(1+p)(1+k)}\,\rho \right).
\end{multline}

\begin{figure}[htbp]
\centering  
\includegraphics[width=0.45\textwidth]{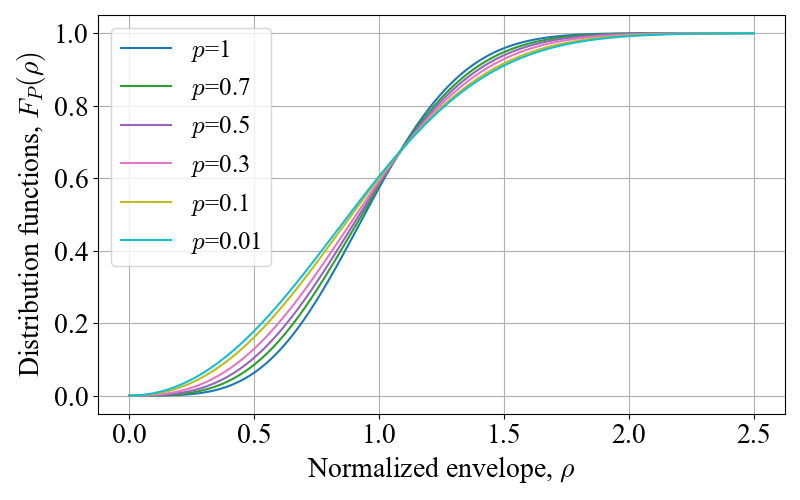}
\caption{The distribution function $F_P(\rho)$ of the extended $k$-$u$ model with different $p$. (As $p$ increases, the proportion of signal amplitudes located at both lower and higher values decreases, leading to an improvement in the fading condition. )}
\label{cdfp}
\end{figure}

Fig.~\ref{cdfp} illustrates the variation of $F_P(\rho)$ with respect to the imbalance parameter $p$ when $k=1$ and $u=2$. As observed, as $p$ gradually increases, the proportion at both ends of the curve decreases, implying a reduction in the probability of deep fading. Meanwhile, the slope of the curve around $\rho=1$ gradually increases, indicating that the envelope becomes more concentrated around $\rho=1$ and the fading condition improves. This behavior is consistent with that of the PDF.

\subsection{Moments of Arbitrary Order}
The expression for the j-th moment of the normalized envelope $P$, denoted by $\mathbb{E}(P^j)$, can be obtained through the following integral:
\begin{align}
\label{35}
E(P^j) &= \int_0^\infty \rho^j f_P(\rho) \mathrm{d}\rho 
= \frac{(1+k)^{\frac{u(1+p)+2}{4}} u(1+p)}{k^{\frac{u(1+p)-2}{4}} e^{\frac{(1+p)uk}{2}}} \notag \\
& \quad \times \int_0^\infty e^{-\frac{(1+k)(1+p)u \rho^2}{2}} \rho^{\frac{u(1+p)+2j}{2}} \notag \\
& \quad \times I_{\frac{u(1+p)-2}{2}}\!\bigl(\sqrt{k(1+k)} u(1+p) \rho\bigr) \mathrm{d}\rho.
\end{align}

We use the following integral formula \cite[eq. (2.15.5.4)]{Integrals and series: Special Functions} to solve the integral in (\ref{35}):
\begin{align}
\label{36}
\int_0^\infty e^{-px^2}x^{a-1}I_v(cx)\,\mathrm{d}x 
= &\frac{\bigl(\frac{c}{2}\bigr)^v\Gamma\bigl(\frac{a+v}{2}\bigr)}{2p^{\frac{a+v}{2}}\Gamma(v+1)} \notag \\
&\times {}_1F_1\biggl(\frac{a+v}{2};v+1;\frac{c^2}{4p}\biggr),
\end{align}
where $\Gamma(x)$ is the Gamma function~~\cite[eq. (6.1.1)]{b29}, and ${}_1F_1(x; y; z)$ is the Kummer hypergeometric function~\cite[eq. (13.1.2)]{b29}. 

The convergence conditions for this integral are: $\mathrm{Re}(a+v)>0,p\in\mathbb{R},|\arg\mathrm{~c}|<\pi.$

By matching the parameters in (\ref{35}) with those in (\ref{36}), the following relationships are obtained:
\begin{equation}
\label{37}
\mathrm{Re}(a+v)=u(1+p)>0,
\end{equation}
\begin{equation}
\label{38}
p=\frac{(1+k)(1+p)u}{2}\in\mathbb{R},
\end{equation}
\begin{equation}
\label{39}
|\arg\mathrm{~c}|=\left|\arg\left(\sqrt{k(1+k)}(1+p)u\right)\right|=0<\pi.
\end{equation}

It can be verified that these relationships satisfy the convergence conditions of (\ref{36}). Therefore, leveraging the integral formula in (\ref{36}) and matching the parameters of (\ref{35}) to it, the integral in (\ref{35}) can be solved. After rearranging the resulting terms, the closed-form expression of $\mathbb{E}(P^j)$ can be obtained as
\begin{align}
\label{j-moment}
\mathbb{E}[P^j] =& \frac{e^{-\frac{(1+p)uk}{2}} \Gamma\left(\frac{u(1+p)+j}{2}\right)}
               {\left(\frac{(1+k)(1+p)u}{2}\right)^{j/2} \Gamma\left(\frac{u(1+p)}{2}\right)} \notag \\
                &\times {}_1F_1 \left( \frac{u(1+p)+j}{2}; \frac{u(1+p)}{2}; \frac{(1+p)uk}{2} \right).
\end{align}

The expression of the moment of arbitrary order of the normalized envelope for the extended $k$–$u$ model is mathematically straightforward, consisting of exponential functions, Gamma functions, and the Kummer confluent hypergeometric function. For a given order $j$, the expression can be further simplified by exploiting the properties of the Gamma and Kummer hypergeometric functions, as demonstrated in the derivation of the amount of fading presented in Section \uppercase\expandafter{\romannumeral 4}.

\section{Performance Evaluation of Wireless Communication Systems}
Based on the fundamental statistics derived in the preceding section, this section conducts the derivation and analysis of several key performance metrics in wireless communication systems, including the amount of fading, outage probability, average bit error rate, and effective rate.
\subsection{Amount of Fading}
The amount of fading $AF$ is an essential metric in wireless communications that quantifies the random fluctuation of channel signals \cite{b32}. It provides a concise measure of the severity of fading and reflects its overall impact on signal transmission. A smaller $AF$ indicates smoother signal power variations and a more stable channel. This metric also guides system design and optimization, including modulation selection, antenna configuration, and channel estimation.

To derive the expression of $AF$ for the extended $k$–$u$ model, the second-order and the fourth-order moments of the normalized envelope must first be obtained. 

The second order moment can be directly obtained from the following equation:
\begin{equation}
\label{2-moment}
\mathbb{E}[P^2] = \mathbb{E}\left[\left(\frac{R}{\sqrt{\mathbb{E}[R^2]}}\right)^2\right] = \frac{\mathbb{E}[R^2]}{\mathbb{E}[R^2]} = 1.
\end{equation}

By setting $j=4$ in (\ref{j-moment}), the fourth order moment of the normalized envelope can be obtained as
\begin{align}
\label{4-moment}
\mathbb{E}[P^4] = &\frac{e^{-\frac{(1+p)uk}{2}} \left(1+\frac{2}{u(1+p)}\right)}{(1+k)^2} \notag \\
& \times{}_1F_1 \left( \frac{u(1+p)}{2}+2; \frac{u(1+p)}{2}; \frac{(1+p)uk}{2} \right).
\end{align}

The Kummer hypergeometric function in (\ref{4-moment}) can be expanded in a series using \cite[eq. (13.1.2)]{b29}:
\begin{equation}
\label{kummer series}
{}_1F_1(a; b; c) = \sum_{N=0}^{\infty} \frac{(a)_N}{(b)_N} \frac{c^N}{N!},
\end{equation}
where $(x)_N$ is the Pochhammer symbol~\cite[Appendix~II.2]{b33}.

After rearrangement, the Kummer hypergeometric function in (\ref{4-moment}) can be simplified to
\begin{align}
\label{17}
{}_1F_1&\left(\frac{u(1+p)}2+2;\ \frac{u(1+p)}2;\ \frac{(1+p)uk}2\right) \notag \\
&= \frac{u(1+p)(1+k)^2+2(1+2k)}{u(1+p)+2}  e^{\frac{(1+p)uk}2}.
\end{align}

By substituting (\ref{17}) back into (\ref{4-moment}) and simplifying, we can obtain the expression for the fourth-order moment as
\begin{equation}
\mathbb{E}[P^4] = 1 + \frac{2(1 + 2k)}{u(1 + p)(1 + k)^2}
\label{4-moment-f}
\end{equation}

By substituting (\ref{2-moment}), and (\ref{4-moment-f}) into the calculation formula fo the amount of fading, i.e., $AF \triangleq \frac{\mathbb{E}[P^4] - \mathbb{E}^2[P^2]}{\mathbb{E}^2[P^2]}$, the final expression for $AF$ can be obtained as
\begin{equation}
\label{18}
AF = \frac{2(1+2k)}{u(1+p)(1+k)^2}.
\end{equation}

The expression of the $AF$ is mathematically relatively simple, which facilitates its application in the design and optimization of wireless communication systems. Meanwhile, the influence of parameter $p$ on the $AF$ can also be directly observed. As parameter $p$ gradually decreases, the $AF$ increases. This implies that the channel fading becomes more severe, and the risk of random interference encountered during signal transmission is higher. 

\subsection{Outage Probability}
The outage probability $P_o$ is defined as the probability that the received signal-to-noise ratio (SNR) $\mathit{\Upsilon}$ falls below a certain reception threshold $\psi$~\cite{b32}. The expression for the outage probability $P_o$ is given by
\begin{equation}
\label{19}
P_o(\psi) = P(\mathit{\Upsilon} < \psi) = F_\mathit{\Upsilon}(\psi).
\end{equation}

To derive the expression of $P_o$ for the extended $k$-$u$ model, it is first necessary to obtain the PDF of the SNR  $\mathit{\mathit{\Upsilon}}$. The SNR  $\mathit{\mathit{\Upsilon}}$  is defined as  $\mathit{\Upsilon} = Z/N_0$, where $Z$  is the signal power and $N_0$ is the noise power. In theoretical analysis, $N_0$ is often set to 1 to simplify the model, so that  $\mathit{\Upsilon} = Z = R^2$. The mean SNR is $\bar{\gamma} = \bar{z} = \mathbb{E}[Z] = \mathbb{E}[R^2] = \hat{r}^2$. Based on the relationship between $\mathit{\mathit{\Upsilon}}$ and $R$, as well as the PDF of $R$ presented in (\ref{rpdf}), the PDF of  $\mathit{\mathit{\Upsilon}}$ denoted by $f_\mathit{\Upsilon}(\gamma)$ can be obtained as
\begin{align}
\label{snrpdf}
f_\mathit{\Upsilon}(\gamma) = &\frac{(\frac{1+k}{\bar{\gamma}})^{\frac{u(1+p)+2}{4}} (1+p)u}{2({\frac{k}{\gamma}})^{\frac{u(1+p)-2}{4}} e^{\frac{(1+p)uk}{2}} } e^{-\frac{(1+k)(1+p)u\gamma}{2\bar{\gamma}}} \notag \\
&\times  I_{\frac{u(1+p)-2}{2}} \left( u(1+p) \sqrt{k(1+k) \frac{\gamma}{\bar{\gamma}}} \right).
\end{align}

Therefore, the outage probability $P_o$ can be obtained as
\begin{align}
\label{21}
P_o(\psi) &= F_\mathit{\Upsilon}(\psi) = \int_0^\psi f_\mathit{\Upsilon}(x)  \mathrm{d}x \notag \\
&= 1 - Q_{\frac{u(1+p)}{2}} \left( \sqrt{(1+p)uk},\ \right. \notag \\
&\quad \left. \sqrt{\frac{(1+k)(1+p)u}{\bar{\gamma}} \psi} \right).
\end{align}

The specific integration process in (\ref{21}) is provided in Appendix A.

Fig.~\ref{Pop} illustrates the outage probability when the average SNR $\bar{\gamma} = 5$. The other parameters are set as follows: $k = 1$, $u = 2$. The variation trend of each curve shows that the outage probability increases with the rise of the reception threshold. When the required outage probability is lower than $1\%$, the reception threshold should preferably be set below $-10$ dB; when the target outage probability is $0.1\%$, the reception threshold needs to be set below $-20$ dB. Taking any curve in the figure as an example, the specific relationship between the outage probability and the reception threshold under the corresponding condition can be analyzed. For instance, for the curve corresponding to $p = 0.75$, the outage probability when the reception threshold is set to $-10$ dB is $86.9\%$ lower than that when the threshold is $-5$ dB; additionally, the outage probability at a reception threshold of $-15$ dB is $85.6\%$ lower than that at $-10$ dB. In other words, under this condition, each $5$ dB decrease in the reception threshold results in an approximately $85\%$ reduction in the outage probability. Furthermore, for a given reception threshold, the outage probability increases as the parameter $p$ decreases. This is because a smaller $p$ leads to more severe fading, which increases the proportion of weak signals and consequently raises the outage probability.

\begin{figure}[htbp]
\centering  
\includegraphics[width=0.45\textwidth]{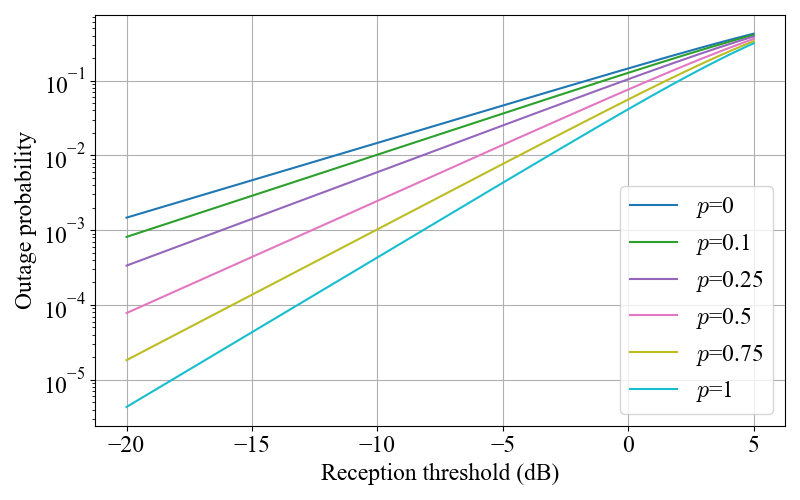}
\caption{Outage probability with different values of $p$ ($k = 1$, $u = 2$).}
\label{Pop}
\end{figure}
\subsection{Average Bit Error Rate}

The ABER of a coherent binary system can be calculated using the following expression ~\cite[eq. (5.3)]{b32}:
\begin{equation}
\label{Pb}
P_b = \frac{1}{\pi} \int_0^{\pi/2} M_\mathit{\Upsilon} \left( \frac{g}{\sin^2 \theta} \right) \mathrm{d}\theta,
\end{equation}
where $g = 1$, $0.5$, $0.715$ correspond to binary phase shift keying (BPSK), binary frequency shift keying (BFSK), and minimum correlation binary phase shift keying (MC-BPSK), respectively. $M_\mathit{\Upsilon}(t)$ is the MGF of $\mathit{\mathit{\Upsilon}}$, and it is defined as
\begin{equation}
\label{mgf}
M_\mathit{\Upsilon}(t) \triangleq \mathbb{E}\{e^{-\mathit{\Upsilon} t}\} = \int_0^\infty e^{-\gamma t} f_\mathit{\Upsilon}(\gamma)  \mathrm{d}\gamma.
\end{equation}

By substituting (\ref{snrpdf}) into (\ref{mgf}), the expression of  $M_\mathit{\Upsilon}(t)$ can be obtained as
\begin{equation}
\label{24}
M_{\mathit{\Upsilon}}(t) = \frac{ e^{ -\frac{uk\overline{\gamma}t(1+p)}{u(1+p)(1+k)+2\overline{\gamma}t} } }
{ \left(1 + \frac{2\overline{\gamma}t}{u(1+p)(1+k)}\right)^{\frac{u(1+p)}{2}} }.
\end{equation}

By substituting (\ref{24}) into (\ref{Pb}), the expression of  $P_b$ can be obtained as
\begin{align}
\label{Pbresult1}
P_b = &\frac{(A/G)^C B(0.5,C+0.5)}{2\pi} \notag \\
&\times  \sum_{n=0}^{\infty} \frac{\left(\frac{-2DA^2}{G}\right)^n {}_2F_1 (C+n,C+0.5;C+1;-A/G)}{n!},
\end{align}
where $A = u(1+p)(1+k)$, $G = 2\bar{\gamma}g$, $C = u(1+p)/2$, and $D = 2uk\bar{\gamma}g(1+p)$, $B(x,y)$ denotes the Beta function~\cite[eq. (6.2.1)]{b29}, and ${}_2F_1(x; y; z)$ denotes the Gauss hypergeometric function~\cite[eq. (7.2.1.1)]{b33}.

When parameters $A$ and $G$ satisfy the condition $|A/G| < 1$, the following series expansion ~\cite[eq. (7.2.1.1)]{b33} of the Gauss hypergeometric function in (\ref{Pbresult1}) meets its convergence condition:
\begin{equation}
\label{26}
_2F_1(a,b;c;z)=\sum_{m=0}^\infty\frac{(a)_m(b)_m}{(c)_m}\frac{z^m}{m!}.
\end{equation}

Substituting (\ref{26}) into (\ref{Pbresult1}), the expression of $P_b$ can be obtained as
\begin{align}
P_b &= \frac{(A/G)^C B(0.5, C+0.5)}{2\pi} \notag \\
&\quad \times \Psi_1 \left( C, C+0.5; C, C+1; -A/G, \frac{-2DA^2}{G} \right),
\label{27}
\end{align}
where $\Psi_1(a, b; c, d; x, y)$ denotes the Horn's hypergeometric series function~\cite[eq. (7.2.4.8)]{b33}.

The derivation process of (\ref{24}), (\ref{Pbresult1}), and (\ref{27}) is presented in Appendix B.

Fig. \ref{ABERp} depicts the ABER under the BPSK modulation scheme. The other parameters are configured as follows: $k=3$, $u=4$, $g=1$. As observed from all curves, ABER decreases with an increase in SNR. For a fixed $p$ value, the specific relationship between ABER and SNR under the corresponding condition can be analyzed by selecting any curve in the figure. For a given SNR, as $p$ increases from 0 to 1, ABER gradually decreases and fading conditions improve. The figure further shows that to achieve an ABER of $10^{-3}$, the SNR should preferably exceed 10 dB; to achieve an ABER of $10^{-4}$, the SNR should preferably exceed 14 dB.
\begin{figure}[htbp]
\centering  
\includegraphics[width=0.5\textwidth]{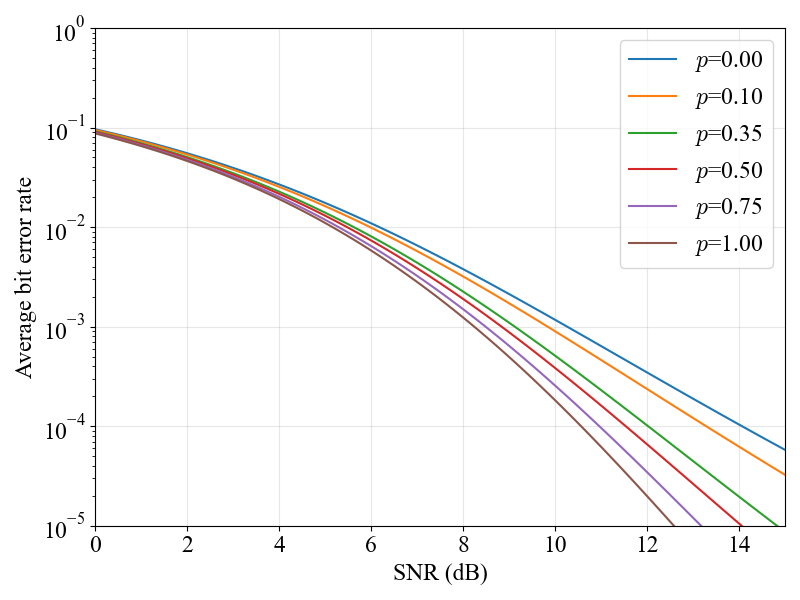}
\caption{ABER under BPSK with different values of $p$ ($k=3$, $u=4$).}
\label{ABERp}
\end{figure}

\subsection{Effective Rate}
The effective rate  is an important metric for evaluating the actual transmission capability of wireless communication systems under quality-of-service (QoS) constraints, such as transmission delay and block duration [34]. It is defined as
\begin{equation}
\label{28}
\mathcal{R}=-\frac{1}{A} \log_2 \left(\int_0^\infty(1+\gamma)^{-A}f_\mathit{\Upsilon}(\gamma)\mathrm{d}\gamma\right),
\end{equation}
where $A\triangleq\frac{\theta TB_c}{\ln2}$, $\theta$ denotes the QoS exponent, $T$ is the block duration, and $B_c$ represents the instantaneous channel capacity.

Substituting (\ref{snrpdf}) into (\ref{28}), the expression of $\mathcal{R}$ can be obtained as
\begin{align}
\label{29}
\mathcal{R} &= \frac{u(1+p)}{-2A}\log_2\!\left(\frac{2\sqrt{\bar{\gamma}}}{u(1+k)(1+p)}\right) \notag \\
&\quad -\frac{1}{A}\left(
    \log_2\!\left(\sum_{m=0}^\infty a_m\right)
    - \log_2\!\left(\overline{\gamma}e^{\tfrac{(1+p)uk}{2}}\right)
\right),
\end{align}
where the expression of $a_m$ is
\begin{equation}
\label{30}
a_m=\frac{E^m}{m!}\times U(m+F;m+1-A+F;H),
\end{equation}
with $E=\frac{k(1+k)u^2(1+p)^2}{4\bar{\gamma}}$, $F=\frac{u(1+p)}{2}$, $H=\frac{(k+1)(1+p)u}{2\bar{\gamma}}$, and $U(a_1;a_2;z)$ is the Tricomi confluent hypergeometric function \cite[eq. (13.1.3)]{b29}. 

The procedure of (\ref{29}) is shown in Appendix C.

\section{Application in Millimeter Wave Bands}
\begin{figure}[htbp]
\centering  
\includegraphics[width=0.5\textwidth, trim=0 0 0 40, clip]{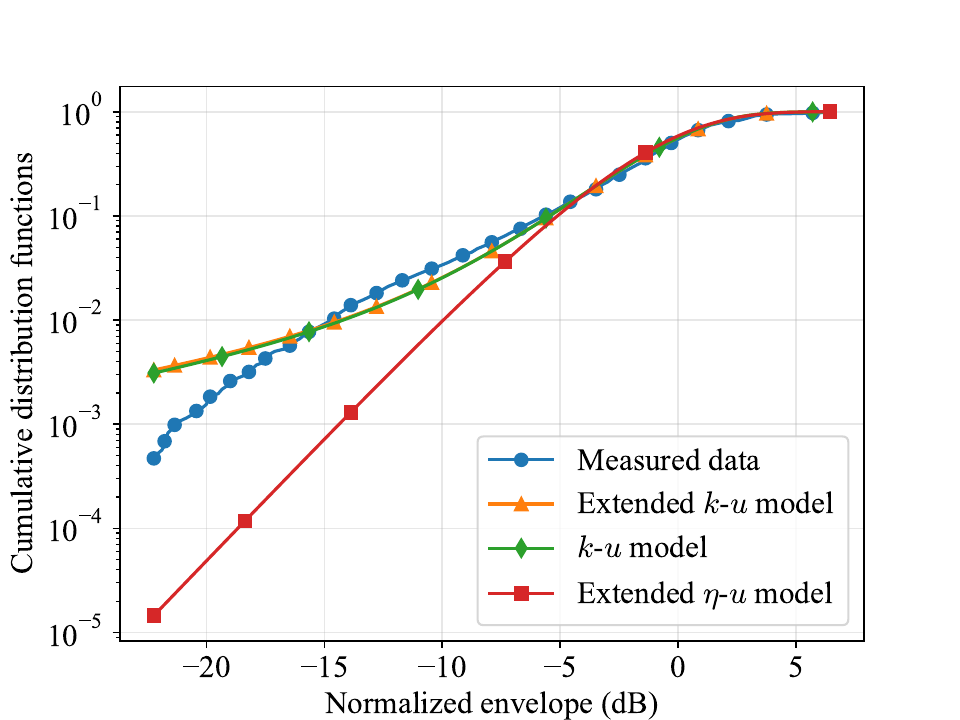}
\caption{Fitting results of different models in 28 GHz LoS Scenario.}
\label{28hz}
\end{figure}

\begin{figure}[htbp]
\centering  
\includegraphics[width=0.5\textwidth, trim=0 0 0 40, clip]{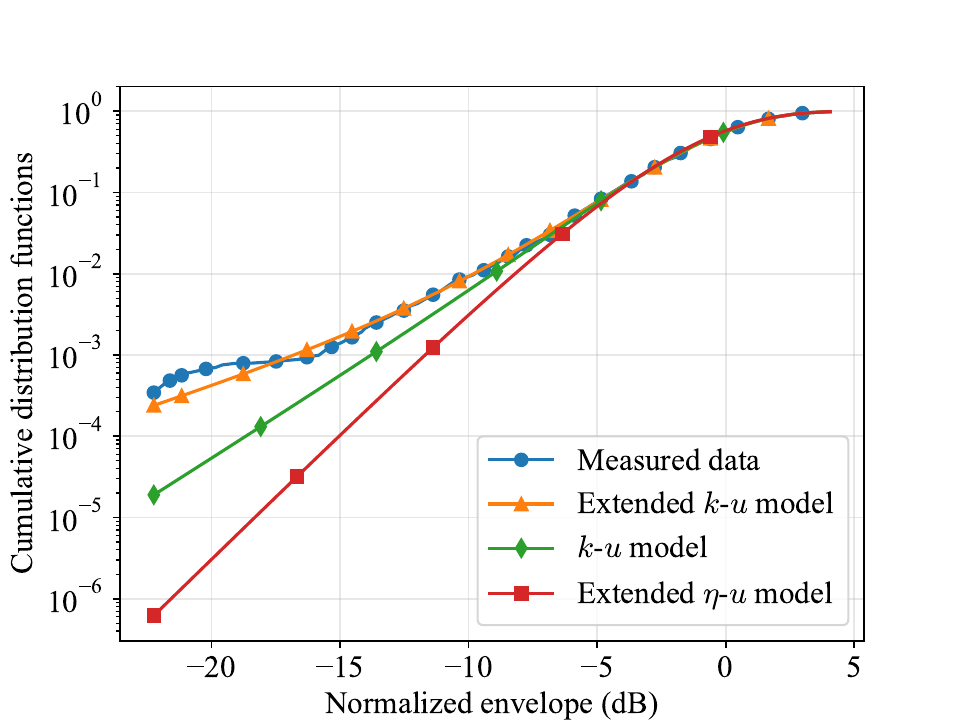}
\caption{Fitting results of different models in 65 GHz LoS Scenario.}
\label{65hz}
\end{figure}

\begin{figure}[htbp]
\centering  
\includegraphics[width=0.5\textwidth, trim=0 0 0 40, clip]{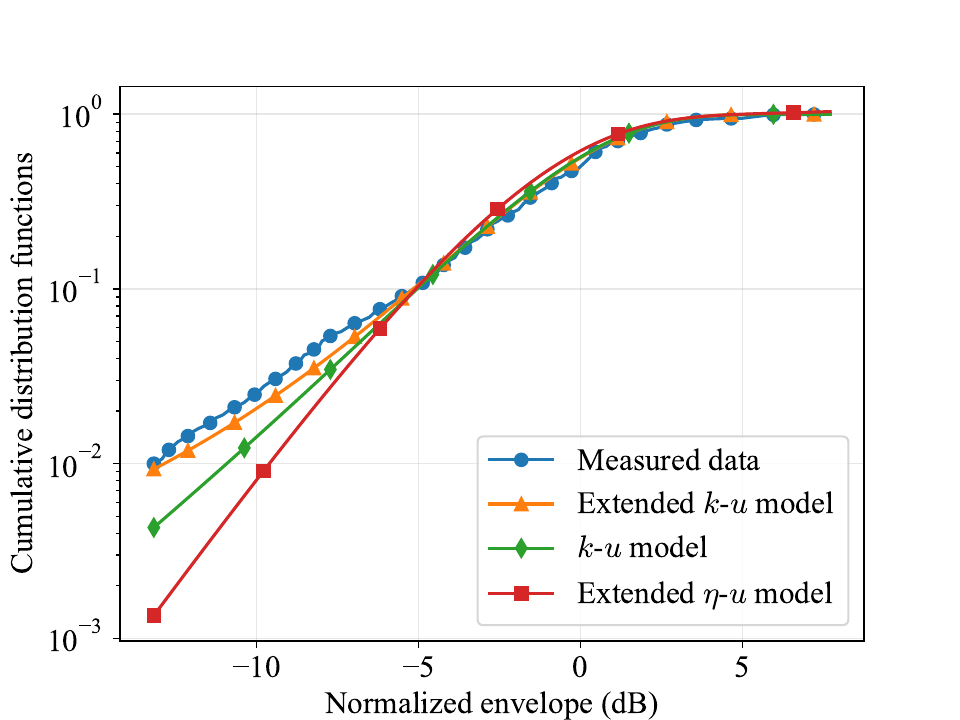}
\caption{Fitting results of different models in 92.5645 GHz LoS Scenario.}
\label{92hz}
\end{figure}

We investigate the application of the extended $k$-$u$ model in mmWave LoS scenarios, and compare it with $k$-$u$ model and extended $\eta$-$u$ model.  Results are presented in Figs.~\ref{28hz}--\ref{92hz}. The measured data in Fig.~\ref{28hz} are extracted from the LoS curve in \cite[Fig. 6]{b35}; these data represent the cumulative distribution of normalized envelope of small-scale fading measured in an outdoor 28 GHz LoS environment. We use the nonlinear least squares function $curvefit$ in Python to fit the extracted data with the cumulative distribution functions of normalized signal envelope  of extended $k$-$u$ model, $k$-$u$ model, and extended $\eta$-$u$ model, resulting in the three fitting curves shown in the figure. It can be observed that as extended $\eta$-$u$ model exhibits less optimal fitting performance in this scenario, which is attributed to the presence of the LoS component. In contrast, the results of extended $k$-$u$ model and $k$-$u$ model are close, and both achieve good fitting performance in this scenario. The measured data in Fig.~\ref{65hz} are extracted from  \cite[Fig.~4(b)]{b4}. This dataset corresponds to fading in an indoor 65 GHz LoS environment. It can be observed that as the frequency increases, the $k$-$u$ model begin to exhibit less satisfactory fitting performance. In contrast, the extended $k$-$u$ model remains capable of effectively characterizing the fading in this scenario. This demonstrates that the introduced imbalance of multipath clusters enhances the characterization capability of extended $k$-$u$ model in mmWave. The measured data in Fig.~\ref{92hz} are extracted from the Experimental curve in \cite[Fig.~5]{b36}, representing fading in an indoor 92.5645 GHz LoS environment. Results show that the extended $k$-$u$ model achieves the best fit to the measured data.

All three experiments are conducted under mmWave LoS scenarios. From these three experiments, it can be observed that the extended $k$-$u$ model performs relatively well across all such mmWave LoS scenarios. In 5G and future wireless communications, signal frequencies will continue to expand toward the mmWave band and even higher frequency bands. The reduction in the number of multipath clusters caused by increased frequencies leads to a growing reliance of communication technologies on LoS components. The extended $k$-$u$ model, which performs well in mmWave scenarios with LoS, can provide a useful tool for the design and analysis of emerging communication systems. 

\section{Conclusion}
In this paper, we propose an extended $k$–$u$ fading model to characterize the imbalance among multipath clusters by introducing an additional parameter $p$ into the original $k$–$u$ model. Notably, when $p = 1$, the proposed model reduces to the original $k$–$u$ model. We derive several key statistical characteristics of the extended $k$–$u$ model, including the probability density function, cumulative distribution function, and moments of arbitrary order. All these statistics are expressed in closed-form expressions, which facilitate the analysis and design of wireless communication systems. This demonstrates that the extended $k$–$u$ model is mathematically more tractable than the $a$–$k$–$\eta$–$u$ model. By examining the variation of the PDF under different values of $p$, we show that the extended $k$–$u$ model can capture a wider range of fading conditions, including more novel and more severe fading scenarios. Based on the derived statistical properties, we further obtain analytical expressions for several essential system performance metrics, such as the amount of fading, outage probability, average bit error rate, and effective rate. To validate the proposed model, we extract fading data reported in previous studies and compare the fitting capability of the extended $k$–$u$ model with that of the original $k$–$s$ model and the extended $\eta$–$u$ model in mmWave LoS scenarios. The results indicate that among the three, the extended $k$–$s$ model provides the best characterization for mmWave LoS conditions, thereby offering a more effective channel modeling tool for 5G and future wireless communication systems with dominant LoS components.

Furthermore, the introduction of the parameter $p$ significantly enhances the flexibility of the model. This flexibility makes the extended $k$–$u$ model particularly attractive for other emerging LoS communication scenarios, such as UAV communications, ground-to-air links, and next-generation wireless networks. It is also worth noting that the extended $\eta$–$u$ model remains well-suited for NLoS mmWave communication environments. Together, the extended $k$–$u$ and $\eta$–$u$ models form a comprehensive modeling framework capable of representing a broad spectrum of mmWave propagation scenarios.

Finally, regarding future research directions, we note that the extended $a$–$\eta$–$u$ model incorporates environmental inhomogeneity on the basis of the extended $\eta$–$u$ model, whereas such inhomogeneity has not yet been considered in the proposed extended $k$–$u$ model. Therefore, future work may focus on integrating environmental inhomogeneity into the extended $k$–$u$ framework to establish a more comprehensive and realistic fading model.
\appendices

\section{}
In Appendix A, we present the detailed integration process in (\ref{21}) for deriving the outage probability $P_o$.

Substituting (\ref{snrpdf}) into (\ref{21}) and rearranging, we obtain the following result:
\begin{align}
\label{40}
F_{\mathit{\Upsilon}}(\gamma) &= \frac{(1+k)^{\frac{u(1+p)+2}{4}} u(1+p)}
{2k^{\frac{u(1+p)-2}{4}} e^{\frac{(1+p)uk}{2}} \overline{\gamma}^{\frac{u(1+p)+2}{4}}} \notag \\
&\quad \times \int_0^\gamma e^{-\frac{(1+k)(1+p)u x}{2\overline{\gamma}}} x^{\frac{u(1+p)-2}{4}} \notag \\
&\quad \times I_{\frac{u(1+p)-2}{2}}\!\left(u(1+p) \sqrt{\frac{k(1+k)}{\overline{\gamma}}} x^{\frac{1}{2}}\right) \mathrm{d}x.
\end{align}

To simplify the expression, we perform the following variable substitutions: $a=\frac{(1+k)^{\frac{u(1+p)+2}{4}}(1+p)u}{2k^{\frac{u(1+p)-2}{4}}e^{\frac{(1+p)uk}{2}}\overline{\gamma}^{\frac{u(1+p)+2}{4}}},$, 	$b=\frac{(1+k)(1+p)u}{2\overline{\gamma}},v=\frac{u(1+p)}{2},c=u(1+p)\sqrt{\frac{k(1+k)}{\bar{\gamma}}}$. With these substitutions, (\ref{40}) can thus be simplified to
\begin{equation}
\label{41}
F_\mathit{\Upsilon}(\gamma)=a\int_0^\gamma e^{-bx}x^{\frac{v-1}{2}}I_{v-1}\left(cx^{\frac{1}{2}}\right).
\end{equation}

We perform another variable substitution by letting $x=\frac{t^2}{2b}$. After rearrangement, (\ref{41}) becomes
\begin{equation}
\label{42}
F_\mathit{\Upsilon}(\gamma)=\frac{2a}{\left(\sqrt{2b}\right)^{v+1}}\int_0^{\sqrt{2b\gamma}}e^{-\frac{t^2}{2}}t^vI_{v-1}\left(\frac{c}{\sqrt{2b}}t\right)\mathrm{d}t.
\end{equation}

To align with the Marcum $Q$ function, (\ref{42}) is rearranged into the following form:
\begin{align}
\label{43}
F_{\Upsilon}(\gamma) = &\frac{2ac^{v-1}}{(2b)^v}e^{\frac{c^2}{4b}} \notag \\
&\times \!\left(1 - \int_{\sqrt{2b\gamma}}^\infty 
\frac{t^v e^{-\frac{t^2}{2}-\frac{c^2}{4b}} I_{v-1}\!\left(\frac{ct}{\sqrt{2b}}\right)}
{\bigl(\frac{c}{\sqrt{2b}}\bigr)^{v-1}} \,\mathrm{d}t\right).
\end{align}
    
Combining with the integral formula of the Marcum $Q$ function in (\ref{34}), (\ref{43}) can be written as
\begin{equation}
\label{44}
F_{\mathit{\Upsilon}}(\gamma)=\frac{2ac^{v-1}}{(2b)^{v}}e^{\frac{c^{2}}{4b}}\left(1-Q_{v}\left(\frac{c}{\sqrt{2b}},\sqrt{2b\gamma}\right)\right).
\end{equation}

Substituting the variable relationships back into (\ref{44}) and rearranging yields the result shown in (\ref{21}).

\section{}
In Appendix B, we present the detailed derivation processes of (\ref{24}), (\ref{Pbresult1}), and (\ref{27}) involved in the ABER calculation.

\subsection{Proof of (\ref{24})}
To derive (\ref{24}), we first substitute (\ref{snrpdf}) into (\ref{mgf}), yielding
\begin{equation}
\label{45}
M_\mathit{\Upsilon}(t)=\frac{(1+k)^{\frac{u(1+p)+2}{4}}(1+p)u}{2k^{\frac{u(1+p)-2}{4}}e^{\frac{(1+p)uk}{2}}\overline{\gamma}^{\frac{u(1+p)+2}{4}}}\times I,
\end{equation}
where the integral $I$ is defined as
\begin{equation}
\label{46}
\begin{aligned}
I &= \int_0^\infty e^{-\left(\frac{(1+k)(1+p)u}{2\overline{\gamma}} + t\right)\gamma} \gamma^{\frac{u(1+p)-2}{4}} \\
  &\quad \times I_{\frac{u(1+p)-2}{2}}\!\left(u(1+p) \sqrt{k(1+k)\frac{\gamma}{\overline{\gamma}}}\right) \mathrm{d}\gamma.
\end{aligned}
\end{equation}

To simplify the expression, we introduce the following variable substitutions:$a=\frac{(1+k)(1+p)u}{2\bar{\gamma}}+t$, $v=\frac{u(1+p)-2}{2}$,$c=u(1+p)\sqrt{k(1+k)\frac{1}{\bar{\gamma}}}$. With these substitutions, (\ref{46}) can be simplified to
\begin{equation}
\label{47}
I=\int_0^\infty e^{-\alpha\gamma}\gamma^{\frac{v}{2}}I_v(c\sqrt{\gamma})\mathrm{~d}\gamma.
\end{equation}

To resolve the square root term in the modified Bessel function of (\ref{47}), we use the series expansion of the modified Bessel function \cite[eq. (9.6.10)]{b29}:
\begin{equation}
\label{48}
I_v(c\sqrt{\gamma})=\sum_{m=0}^\infty\frac{\left(\frac{c}{2}\right)^{2m+v}\gamma^{m+\frac{v}{2}}}{m!\Gamma(m+v+1)}.
\end{equation}

Substituting (\ref{48}) into (\ref{47}) and rearranging terms, we obtain
\begin{equation}
\label{49}
I = \left(\frac{c}{2}\right)^{2m+v} \sum_{m=0}^\infty \frac{\int_0^\infty a^{-m-v-1} e^{-a\gamma} (a\gamma)^{m+v} d(a\gamma)}{m! \Gamma(m+v+1)}.
\end{equation}

The gamma function satisfies the following integral formula \cite[eq. (6.1.1)]{b29}:
\begin{equation}
\label{50}
\Gamma(z)=\int_0^\infty e^{-x}x^{z-1}\mathrm{~d}x\quad(z>0).
\end{equation}

By matching the parameters in (\ref{50}) with (\ref{49}), we obtain
\begin{equation}
\label{51}
z=m+v+1=m+\frac{u(1+p)}{2}>0.
\end{equation}

Thus, we can use (\ref{50}) to evaluate the integral in (\ref{49}). After evaluating the integral and combining the series using the series expansion of the exponential function, (\ref{49}) finally becomes
\begin{equation}
\label{52}
I=\frac{1}{a^{v+1}}\left(\frac{C}{2}\right)^ve^{\frac{c^2}{4a}}.
\end{equation}

Substituting (\ref{52}) back into (\ref{45}), and further substituting the variable relationships back into the resulting expression, we can obtain the result shown in  (\ref{24}) after rearrangement.

\subsection{Proof of (\ref{Pbresult1})}
To derive (\ref{Pbresult1}), we substitute (\ref{24}) into (\ref{Pb}), which gives
\begin{align}
\label{53}
P_b &= \frac{1}{\pi} \int_0^{\frac{\pi}{2}} 
\left( \frac{u(1+p)(1+k)\sin^2\theta}{u(1+p)(1+k)\sin^2\theta + 2\overline{\gamma}g} \right)^{\frac{u(1+p)}{2}} \notag \\
&\quad \times e^{\displaystyle -\frac{uk\overline{\gamma}g(1+p)}{u(1+p)(1+k)\sin^2\theta + 2\overline{\gamma}g}} \mathrm{d}\theta.
\end{align}

To simplify the expression, we introduce the following substitutions of variables: $\bar{A}=u(1+p)(1+k)$, $G=2\bar{\gamma}g$, $C=\frac{u(1+p)}{2}$, and $D=2uk\bar{\gamma}g(1+p)$. With these substitutions, (\ref{53}) can be simplified to
\begin{equation}
\label{54}
P_b=\frac{1}{\pi}\int_0^{\frac{\pi}{2}}\left(\frac{A\sin^2(\theta)}{A\sin^2(\theta)+G}\right)^Ce^{\frac{-D}{2A\sin^2(\theta)+2G}}\mathrm{d}\theta.
\end{equation}

We then perform another substitution for the variable by letting $x=\sin^{2}(\theta)$. After rearrangement, (\ref{54}) can be rewritten as
\begin{equation}
\label{55}
P_b = \frac{1}{2\pi} \int_0^1 \frac{x^{C-0.5}}{\sqrt{1-x}} \left(x+\frac{G}{A}\right)^{-C} e^{-\frac{D}{2(Ax+G)}} \mathrm{d}x.
\end{equation}

Expanding the exponential function in (\ref{55}) as a series and rearranging terms, we get the following expression:
\begin{align}
\label{56}
P_b &= \sum_{n=0}^\infty \left[ \frac{(-2AD)^n}{2\pi n!} \right. \notag \\
&\quad \left. \times \int_0^1 x^{C-0.5} \left(x + \frac{G}{A}\right)^{-C-n} (1-x)^{-1/2} \mathrm{d}x \right].
\end{align}

We use the following integral formula \cite[eq. (3.197.8)]{b30} to evaluate the integral in (\ref{56}):
\begin{align}
\label{57}
&\int_0^b x^{v-1}(x+a)^\lambda(b-x)^{d-1}\,\mathrm{d}x \notag \\
&= a^\lambda b^{d+v-1}B(d,v) \times {}_2F_1\!\bigl(-\lambda,\,v;\,d+v;\,-\tfrac{b}{a}\bigr).
\end{align}

The convergence conditions for this integral are: $|\arg\left(\frac{b}{a}\right)|<\pi,\mathrm{Re}\left(d\right)>0,\mathrm{Re}\left(v\right)>0.$

By matching the parameters in (\ref{57}) with those in (\ref{56}), we obtain the following results:
\begin{equation}
\label{58}
\left|\arg\left(\frac{b}{a}\right)\right|=\left|\arg\left(\frac{u(1+p)(1+k)}{2\bar{\gamma}g}\right)\right|=0<\pi,
\end{equation}
\begin{equation}
\label{59}
\mathrm{Re}\left(d\right)=\mathrm{Re}\left(0.5\right)>0,
\end{equation}
\begin{equation}
\label{60}
\mathrm{Re}\left(v\right)=\mathrm{Re}\left(\frac{u(1+p)}{2}+0.5\right)>0.
\end{equation}

These conditions satisfy the convergence requirements of (\ref{57}). By applying (\ref{57}) to evaluate the integral in (\ref{56}), we can rearrange the resulting expression to obtain the result shown in (\ref{Pbresult1}).

\subsection{Proof of (\ref{27})}
To derive (\ref{27}), we first substitute (\ref{26}) into (\ref{Pbresult1}), yielding
\begin{align}
\label{61}
P_b &= \frac{\left(\frac{A}{G}\right)^C B(0.5, C+0.5)}{2\pi} \notag \\
    &\quad \times \sum_{n=0}^\infty \sum_{m=0}^\infty \left[ \frac{(C+n)_m (C+0.5)_m}{(C+1)_m} \right. \notag \\
    &\quad \left. \times \frac{\left(-\frac{A}{G}\right)^m}{m!} \frac{\left(\frac{-2DA^2}{G}\right)^n}{n!} \right].
\end{align}

Using the property of the Pochhammer symbol [33, Appendix \uppercase\expandafter{\romannumeral 2}.2]:
\begin{equation}
\label{62}
(C)_m=\frac{\Gamma(C+m)}{\Gamma(C)}.
\end{equation}

The term $(C+n)_m$ in (\ref{61}) can be transformed as follows
\begin{align}
\label{63}
(C+n)_m &= \frac{\Gamma(C+n+m)}{\Gamma(C+n)} \notag \\
&= \frac{\Gamma(C+n+m)/\Gamma(C)}{\Gamma(C+n)/\Gamma(C)} = \frac{(C)_{n+m}}{(C)_n}.
\end{align}

Substituting this transformation into (\ref{61}) and rearranging terms, we obtian
\begin{align}
\label{64}
P_b &= \frac{\left(\frac{A}{G}\right)^C B(0.5, C+0.5)}{2\pi} \notag \\
    &\quad \times \sum_{m=0}^\infty \sum_{n=0}^\infty \left[ \frac{(C)_{n+m} (C+0.5)_m}{(C)_n (C+1)_m} \right. \notag \\
    &\quad \left. \times \frac{\left(-\frac{A}{G}\right)^m}{m!} \frac{\left(\frac{-2DA^2}{G}\right)^n}{n!} \right].
\end{align}

Using the following formula \cite[eq. (7.2.4.8)]{b33}, we can directly obtain the result shown in (\ref{27}) from (\ref{64}) as
\begin{equation}
\label{65}
\Psi_1(a,b;c,d;x,y)=\sum_{k=0}^\infty\sum_{l=0}^\infty\frac{(a)_{k+l}(b)_k}{(c)_k(d)_l}\frac{(x)^k}{k!}\frac{(y)^l}{l!}.
\end{equation}

\section{}
In Appendix C, we present the detailed derivation process of (\ref{29}) involved in the effective rate calculation. Substituting  (\ref{snrpdf}) into (\ref{28}), we obtain
\begin{equation}
\label{66}
\mathcal{R}=-\frac{1}{A}log_2\left(\frac{(1+k)^{\frac{u(1+p)+2}{4}}(1+p)u}{2k^{\frac{u(1+p)-2}{4}}e^{\frac{(1+p)uk}{2}}\overline{\gamma}}\times I\right),
\end{equation}
where the integral $I$ is defined as
\begin{align}
\label{67}
I &= \int_0^\infty (1+\gamma)^{-A} e^{-\frac{(1+k)(1+p)u\gamma}{2\overline{\gamma}}} \left(\frac{\gamma}{\overline{\gamma}}\right)^{\frac{u(1+p)-2}{4}} \notag \\
  &\quad \times I_{\frac{u(1+p)-2}{2}}\!\left(u(1+p) \sqrt{k(1+k)\frac{\gamma}{\overline{\gamma}}}\right) \mathrm{d}\gamma.
\end{align}

To simplify the expression, we introduce the following variable substitutions: $a=\frac{(k+1)(1+p)u}{2\bar{\gamma}}$, $v=\frac{u(1+p)}{2}$, and $c=u(1+p)\sqrt{\frac{k(1+k)}{\bar{\gamma}}}$. With these substitutions, (\ref{67}) can be simplified to
\begin{equation}
\label{68}
I=\int_{0}^{\infty}(1+\gamma)^{-A}e^{-\alpha\gamma}\left(\frac{\gamma}{\bar{\gamma}}\right)^{\frac{v-1}{2}}I_{v-1}\left(c\gamma^{\frac{1}{2}}\right)\mathrm{d}\gamma.
\end{equation}

We use the series expansion of the modified Bessel function in (\ref{48}) to (\ref{68}). After substitution and rearrangement,  (\ref{68}) becomes
\begin{align}
\label{69}
I = &\left( \frac{c}{2\sqrt{\overline{\gamma}}} \right)^{v-1} \times \sum_{m=0}^\infty \left[ \frac{\left( \frac{c^2}{4} \right)^m}{m! \Gamma(m+v)} \right. \notag \\
  &\times \left. \int_0^\infty \gamma^{m+v-1} (1+\gamma)^{-A} e^{-a\gamma} d\gamma \right].
\end{align}

We utilize the integral representation of the Tricomi confluent hypergeometric function \cite[eq. (3.197.8)]{b30}:
\begin{equation}
\label{70}
U(a_1;a_2;z)=\frac{1}{\Gamma(a_1)}\int_0^\infty t^{\alpha_1-1}(1+t)^{\alpha_2-\alpha_1-1}e^{-zt}\mathrm{d}t.
\end{equation}

The convergence conditions for this integral are: $\mathrm{Re}(a_1)>0,\mathrm{Re}(z)>0.$

By matching the parameters in (\ref{70}) with those in (\ref{69}), we obtain
\begin{equation}
\label{71}
\mathrm{Re}(a_1)=\mathrm{Re}\left(m+\frac{u(1+p)}{2}\right)>0,
\end{equation}
\begin{equation}
\label{72}
\mathrm{Re(z)}=\mathrm{Re}\left(\frac{(k+1)(1+p)u}{2\overline{\gamma}}\right)>0.
\end{equation}

These conditions satisfy the convergence requirements of (\ref{70}). By applying (\ref{70}) to evaluate the integral in (\ref{69}), we can rearrange the resulting expression to obtain
\begin{align}
\label{73}
I &= \left( \frac{c}{2\sqrt{\overline{\gamma}}} \right)^{v-1} \notag \\
  &\quad \times \sum_{m=0}^\infty \left( \frac{\left( \frac{c^2}{4} \right)^m}{m!} 
    U(m+v; m+v+1-A; a) \right).
\end{align}

Substituting (\ref{73}) back into (\ref{66}), and further substituting the variable relationships back into the resulting expression, we finally obtain the result shown in (\ref{29}) after rearrangement.

\vfill


\begin{thebibliography}{00}
\bibitem{b1} T. S. Rappaport, \textit{Wireless Communications: Principles and Practice}, 2nd ed. Englewood Cliffs, NJ: Prentice-Hall, 2002.
\bibitem{b2} M. Nakagami, “The m-Distribution—A General Formula of Intensity Distribution of Rapid Fading," \textit{Statistical Methods in Radio Wave Propagation}, W. C. Hoffman, Ed. Elmsford, NY: Pergamon, 1960, pp. 3-36.
\bibitem{b3} L. Rayleigh, “On the resultant of a large number of vibrations of the same pitch and of arbitrary phase,” \textit{The London, Edinburgh, and Dublin PhiLoSophical Magazine and Journal of Science}, vol. 10, no. 60, pp. 73–78, 1880.
\bibitem{b4} T. R. R. Marins, A. A. Dos Anjos, V. M. R. Peñarrocha, L. Rubio, J. Reig, R. A. A. de Souza, and M. D. Yacoub, ‘‘Fading evaluation in the mm-wave band,’’ \textit{IEEE Transactions on Communications}., vol. 67, no. 12, pp. 8725–8738, Dec. 2019.
\bibitem{b5} O. S. Badarneh and M. S. Aloqlah, “Performance analysis of digital communication systems over $\alpha$-$\eta$-$\mu$ fading channels,” \textit{IEEE Transactions on Vehicular Technology}., vol. 65, no. 10, pp. 7972–7981, Oct. 2015. 
\bibitem{b6} M. D. Yacoub, “The $\eta$-$\mu$ distribution: A general fading distribution,”  in \textit{Proc. IEEE 52nd Veh. Technol. Conf.-Fall}, vol. 2, 2000, pp. 872–877.
\bibitem{b7} M. D. Yacoub, “The $\kappa$-$\mu$ distribution: A general fading distribution,”  in \textit{Proc. IEEE 54th Veh. Technol. Conf.-Fall}, vol. 3. IEEE, 2001, pp. 1427–1431.
\bibitem{b8} M.D. Yacoub, “The $\kappa$-$\mu$ distribution and the $\eta$-$\mu$ distribution, ” \textit{IEEE Antennas and Propagation Magazine}, vol. 49, pp. 68-81, Feb., 2007.
\bibitem{b9} W. E. Vieira, P. L. L. Bertarini, K. B. Carbonaro, G. A. Carrijo, E. Agustini, and A. A. D. Anjos, “Performance evaluation of cross M-QAM modulation over standardized RF2 5G frequency bands using the $\eta$-$\mu$ fading model,” \textit{IEEE Access}, vol. 13, pp. 31024-31031, Feb., 2025.
\bibitem{b10} B. Karahan and I. Develi, "The Impact of Transceiver IQI on RIS-Assisted OFDM-IM over $\eta$-$\mu$ Fading Channels," in \textit{Proc. IEEE Int. Conf. Commun., Netw. Satell. Syst.}.  IEEE, 2024, pp. 641–645.
\bibitem{b11} P. S. Bithas, V. Nikolaidis, A. G. Kanatas, and G. K. Karagiannidis, “UAV-to-Ground communications: Channel modeling and UAV selection,” \textit{IEEE Transactions on Communications}, vol. 68, no. 8, pp. 5135–5144, Aug., 2020.
\bibitem{b12} S. Thaherbasha and N. P. SD, “Outage performance of IRS-NOMA systems in $\kappa$-$\mu$ fading environment: A multi-user analysis considering imperfect CSI and SIC,” \textit{Wireless Networks}, pp. 1–14, 2025.
\bibitem{b13} V. Saiprudhvi and H. Subramaniyam, “Performance analysis of hybrid OTFS system over $\kappa$-$\mu$ fading channels,” in \textit{Proc. IEEE 14th Int. Conf. Comput. Commun., Control Netw. Technol.}, 2023, pp. 1-5.
\bibitem{b14} M.D. Yacoub, "The $\alpha$-$\mu$ distribution: A general fading distribution," \textit{Proc. IEEE Int. Symp. Pers., Indoor Mobile Radio Commun.}, vol.2, 2002, pp. 629-633.
\bibitem{b15} V. M. Rodrigo-Peñarrocha, J. Reig, L. Rubio, H. Fern´andez, and S. Loredo, “Analysis of small-scale fading distributions in vehicle-to-vehicle communications,” \textit{Mobile Information Systems}, vol. 2016, pp. 1-7, Aug., 2016.
\bibitem{b16} G. Fraidenraich and M. D. Yacoub. “The $\alpha$-$\eta$-$\mu$ and $\alpha$-$\kappa$-$\mu$ fading distributions,” in \textit{Proc. IEEE 9th Int. Symp. Spread Spectr. Tech. Appl.}, 2006, pp. 16-20.
\bibitem{b17} W. Cheng, X. Ma, J. Chen, and G. Nie, “On the unified statistics of cascaded $\alpha$-$\eta$-$\mu$ and $\alpha$-$\kappa$-$\mu$ fading models with applications,” in \textit{IEEE/CIC Int. Conf. Commun. China}, 2024, pp.740-745.
\bibitem{b18} D. Krstic, N. Petrovic, S. Suljovic, G. K. Pandey, D. S. Gurjar, and S. Yadav, “AI-driven approach for QoS estimation using LCR in 5G network with selection combining in $\alpha$-$\eta$-$\mu$ fading and co-channel interference environment,” in \textit{Proc. IEEE Int. Symp. Silicon-on-Insulator Technol. Devices}, 2023, pp.1-6.
\bibitem{b19} J. M. Moualeu, D. B. da Costa, W. Hamouda, U. S. Dias, and R. A. A. de Souza, “Physical layer security over $\alpha$-$\eta$-$\mu$ and $\alpha$-$\kappa$-$\mu$ fading channels,” \textit{IEEE Transactions on Vehicular Technology}, vol. 68, pp. 1025–1029, 2019.
\bibitem{b20} M. Pätzold and G. Rafiq, “Sparse multipath channels: Modelling, analysis, and simulation,”  in \textit{Proc. IEEE 24th Annu. Int. Symp. Pers., Indoor, Mobile Radio Commun.}. IEEE, 2013, pp. 30–35.
\bibitem{b21} A. Hughes, S. Y. Jun, C. Gentile, D. Caudill, J. Chuang, J. Senic, and D. G. Michelson, “Measuring the impact of beamwidth on the correlation distance of 60 ghz indoor and outdoor channels,” \textit{IEEE Open Journal of Vehicular Technology}, vol. 2, pp. 180–193, 2021.
\bibitem{b22} M. D. Yacoub. "The $\alpha$-$\eta$-$\kappa$-$\mu$ Fading Model." \textit{IEEE Transactions on Antennas and Propagation}, vol. 64, no. 8, pp. 3597-3610, Aug., 2016.
\bibitem{b23} G. R. de L. Tejerina, C. R. N. da Silva, and M. D. Yacoub. “Extended $\eta$-$\mu$ fading models,” \textit{IEEE Transactions on Wireless Communications}, vol. 19, no. 12, pp. 8153-8164, Sept., 2020.
\bibitem{b24} H. Al-Hmood, R. S. Abbas, and H. Al-Raweshidy, “Extended $\alpha$-$\eta$-$\mu$ fading distribution: Statistical properties and applications”, \textit{IEEE Access}, vol. 10, pp. 109803-109813, Jan., 2022.
\bibitem{b25} T. R. R. Marins, A. A. Dos Anjos, C. R. N. Da Silva, V. M. R. Peñarrocha, L. Rubio, J. Reig, R. A. A. De Souza, and M. D. Yacoub, “Fading evaluation in standardized 5G millimeter wave band,” \textit{IEEE Access}, vol.9, pp. 67268-67280, Apr., 2021.
\bibitem{b26} M. S. Aloqlah and Q. M. Bashayreh, “Error analysis of binary coherent and M-QAM signaling in extended $\eta$-$\mu$ fading channels with AWGN,” in \textit{Proc. IEEE Middle East Conf. Commun. Netw.}, 2024, pp. 211-216.
\bibitem{b27} G. R. de Lima Tejerina, C. R. N. da Silva, R. A. A. de Souza and M. D. Yacoub, “On the extended $\eta$-$\mu$ model: new results and applications to IRS-Aided systems,” \textit{IEEE Transactions on Vehicular Technology}, vol. 72, no. 4, pp. 4133-4142, Apr., 2022.
\bibitem{b28} Z. Hussain, H. Mehdi, and S. M. Saleem, “Relay-assisted D2D communication over extended $\eta$-$\mu$ fading channels,” \textit{Mobile Information Systems}, 2022.
\bibitem{b29}
M. Abramowitz and I. A. Stegun, \textit{Handbook of Mathematical Functions}, Washington, DC: U.S. Dept. of Commerce, National Bureau of Standards, 1972.

\bibitem{b30}
L. S. Gradshteyn and L. M. Ryzhik, \textit{Table of Integrals, Series, and Products}, 7th ed. New York: Academic Press, 2007.

\bibitem{b31}
A. P. Prudnikov, Y. A. Brychkov, and O. I. Marichev, \textit{Integrals and Series: Inverse Laplace Transforms}, Gordon and Breach Science, vol. 5, 1992.

\bibitem{b32}
M. K. Simon and M.-S. Alouini, \textit{Digital Communication Over Fading Channels: A Unified Approach to Performance Analysis}, New York: Wiley, 2000.
\bibitem{Integrals and series: Special Functions} A. P. Prudnikov, Y. A. Brychkov, and O. I. Marichev, \textit{Integrals and series: Special Functions}, Gordon and Breach Science, vol. 2, 1992.
\bibitem{b33}
A. P. Prudnikov, Y. A. Brychkov, and O. I. Marichev, \textit{Integrals and Series: More Special Functions}, Gordon and Breach Science, vol. 3, 1992.
\bibitem{b34} S. K. Yoo, S. L. Cotton, P. C. Sofotasios, S. Muhaidat, and G. K. Karagiannidis, ‘‘Effective capacity analysis over generalized composite fading channels,’’ \textit{IEEE Access}, vol. 8, pp. 123756–123764, June, 2020.
\bibitem{b35} M. K. Samimi, G. R. MacCartney, S. Sun, and T. S. Rappaport, “28 ghz millimeter wave ultrawideband small-scale fading models in wireless channels,” in \textit{Proc. IEEE 83rd Veh. Technol. Conf.-Spring}. IEEE, 2016, pp. 1–6.
\bibitem{b36} J. Reig, V. M. R. Peñarrocha, L. Rubio, M. T. M. Inglés, and J. M. Molina-García-Pardo, "The folded normal distribution: A new model for the small-scale fading in line-of-sight (LoS) condition," \textit{IEEE Access}, vol. 7, pp. 77328-77339, Jan., 2019.

\end{thebibliography}
\end{document}